\documentclass[conference]{IEEEtran}
\IEEEoverridecommandlockouts
\pdfoutput=1 
\usepackage[utf8]{inputenc}
\usepackage[T1]{fontenc}
\usepackage{amsmath,amssymb,amsfonts}
\usepackage{graphicx}
\usepackage{subcaption}
\usepackage{textcomp}
\usepackage[table,xcdraw]{xcolor}
\usepackage{xcolor}
\usepackage{multirow}
\usepackage{tabularx,ragged2e,booktabs,caption}
\usepackage{biblatex} 
\usepackage{enumitem}
\usepackage{pifont}
\usepackage{quantikz}
\def\BibTeX{{\rm B\kern-.05em{\sc i\kern-.025em b}\kern-.08em
    T\kern-.1667em\lower.7ex\hbox{E}\kern-.125emX}}
\usepackage{xspace}
\usepackage{flushend}
\usepackage{algorithm}
\usepackage{algorithmic}
\usetikzlibrary{shapes,arrows}
\usepackage{tikz}
\usetikzlibrary{positioning}
\usepackage{booktabs} 
\usepackage{array}    

\begin{document}

\newcommand{\HW}[1]{\textcolor{blue}{[HW: #1]}}

\title{Prepare Ansatz for VQE with Diffusion Model}

\author{\IEEEauthorblockN{
Yilin Shen  \ \
\IEEEauthorblockA{
University of Pennsylvania, PA, USA.
\vspace{-0.15in}}
}
}

\maketitle

\begin{abstract}

The Variational Quantum Eigensolver (VQE) is a quantum algorithm used to find the ground state energy of a given Hamiltonian. The key component of VQE is the ansatz, which is a trial wavefunction that the algorithm uses to approximate the ground state. Designing a good ansatz can significantly improve the performance of the VQE algorithm.
Typical ansatz structures include the Unitary Coupled Cluster (UCC) ansatz and the Hardware-Efficient Ansatz (HEA). The primary distinction between these two structures lies in their dependence on the problem and hardware. The UCC ansatz is tailored to the target Hamiltonian, whereas the HEA is determined by the hardware topology. We believe that an intermediate approach could combine the benefits of the UCC ansatz while introducing additional parameters to increase its expressiveness and capability.
In this paper, we propose utilizing a diffusion model to facilitate the generation of ansatz. We create a sequence of UCC ansatzes as training data and input this data into the diffusion model. The model then generates quantum circuits that have a similar structure to the input data. These quantum circuits are subsequently tested using a VQE task to evaluate their performance. This approach provides a systematic method for generating ansatzes that maintain a similar structure while incorporating additional parameters, enhancing their expressiveness and capability.
We validate on small molecules that the diffusion model can help prepare ansatz circuits for VQE. 

\end{abstract}

\begin{IEEEkeywords}
Variational Quantum Circuit, Quantum Computing, Variational Quantum Eigensolver, Diffusion Model
\end{IEEEkeywords}

\section{Introduction}

Quantum computing is a rapidly developing field with the potentials to solve complex problems~\cite{davies2023tools,iniguez2023quantum,epstein2023algorithmic,roy2022realization,yard2022onchip}.
As one of the most promising quantum algorithms, the Variational Quantum Eigensolver (VQE) has proven its effectiveness in simulating molecular behavior. VQE efficiently calculates the ground state energies of molecular systems, which is essential for understanding their properties and interactions~\cite{xing2023hybrid,larsen2023feedbackbased,shee2022quantum}.
In the VQE algorithm, a parameterized quantum circuit (PQC) is employed to approximate the quantum states associated with a molecular system. During the training process, the parameters within the PQC are iteratively updated to minimize the expectation values, which correspond to the ground state energy of the molecule. This technique allows for accurate and efficient approximation of the molecular system's properties and behavior~\cite{gunlycke2023cascaded,endo2023optimal,fitzpatrick2022selfconsistent}.
Consequently, designing an appropriate parameterized quantum circuit, also known as an ansatz, is crucial for improving the performance of the VQE algorithm.

The Unitary Coupled Cluster (UCC)~\cite{romero2018strategies} and Hardware-Efficient (HEA)~\cite{kandala2017hardware} ansatz are commonly used structures in quantum computing. UCC incorporates physical information of the molecule, while HEA considers the hardware topology. UCC requires a large number of layers and gates, making it inefficient, while HEA ignores the physical information and can lead to suboptimal results.
The Hardware-Efficient Ansatz (HEA) employs single-qubit parameterized gates on all qubits and two-qubit parameterized gates on all possible connections, with most of the gates being parameterized. On the other hand, the Unitary Coupled Cluster (UCC) Ansatz uses trotterization to simulate the exponential of Hamiltonian on qubits to approximate the states of the molecular system. The UCC Ansatz involves a lower proportion of parameterized gates compared to the HEA Ansatz.
Recently, researchers propose to adopt Neural Architecture Search (NAS)~\cite{wang2022quantumnas} to search for more efficient ansatz structures. 
The proposed NAS method starts from multi-layer HEA as super-circuits, it samples sub-circuits and uses evolutionary algorithm to search for the better ansatz structure.

This paper introduces a machine learning-based approach to create ansatz structures for VQE algorithms. 
Incorporation of the physical information of the target molecule system brings the advantages of the ``gold standard'' ansatz UCC over HEA. At the same time, UCC has fewer parameterized gates than HEA. Therefore, we propose to boost the flexibility of the current UCC by inserting some parameterized gates. We propose to use a diffusion model to generate ansatz structures that have a similar structure to UCC. Initially, we generate a set of UCC ansatz and encode them into images, which are then fed into the diffusion model. Once we obtain the generated images from the diffusion model, we decode them into quantum circuits and evaluate their performance on VQE tasks. The key insight is to use diffusion model to generate images that preserve the structures of the original images.

We validate that the generated ansatz works for small molecules including $H_2,\ LiH\ and\ H_2O$. On these VQE tasks, the generated ansatz demonstrates superiror performance over randomly generated ansatz.

\section{Background}

\subsection{VQE and ansatz circuit}

The variational quantum eigensolver (VQE) uses hybrid quantum-classical computation to calculate eigenvalues of Hamiltonians. VQE has demonstrated its efficiency in solving the electronic Schrödinger equation for various small molecules. However, the performance of VQE largely depends on the selection of the variational ansatz that is used to represent the trial wave function. Therefore, constructing an effective ansatz is an active field of research.
Once the parameterized circuit or the ansatz is generated, the ansatz parameters are then iteratively updated in a variational approach until the expectation value of the electronic Hamiltonian is minimized.
The Hamiltonian of a quantum system can be described as:
\begin{equation}
H = \sum_{i} h_i P_i
\end{equation}
where $H$ is the Hamiltonian, $h_i$ are coefficients, and $P_i$ are the Pauli operators.
Then ansatz circuit is adopted to generate a prepared quantum state:
\begin{equation}
|\psi(\vec{\theta})\rangle = U(\vec{\theta})|0\rangle
\end{equation}
where $\vec{\theta}$ is a vector of parameters, $U(\vec{\theta})$ is a parameterized unitary operation, and $|0\rangle$ is the initial quantum state. We can obtain the expectation value of the Hamiltonian:
\begin{equation}
\langle H(\vec{\theta})\rangle = \langle\psi(\vec{\theta})|H|\psi(\vec{\theta})\rangle
\end{equation}
The expectation value (ground state energy) is our objective function to minimize:
\begin{equation}
E(\vec{\theta}) = \min_{\vec{\theta}} \langle H(\vec{\theta})\rangle
\end{equation}
In each iteration, the parameters are updated according to the optimization algorithm:
\begin{equation}
\vec{\theta}_{k+1} = \vec{\theta}_k - \eta \nabla E(\vec{\theta}_k)
\end{equation}
where $\vec{\theta}_k$ is the parameter vector at iteration $k$, $\eta$ is the learning rate, and $\nabla E(\vec{\theta}_k)$ is the gradient of the objective function with respect to the parameters at iteration $k$.

The molecular Hamiltonian in its electronic structure form:
\begin{equation}
\hat{H} = \hat{H}_{1} + \hat{H}_{2}
\end{equation}
where $\hat{H}_{1}$ represents the one-electron terms , and $\hat{H}_{2}$ represents the two-electron terms.
\begin{align}
  \hat{H}_{1} &= \sum_{p,q}{h_{pq} \hat{a}_p^\dagger \hat{a}_q} \\
  \hat{H}_{2} &= \frac{1}{2} \sum_{p,q,r,s} {h_{pqrs} \hat{a}_p^\dagger \hat{a}_q^\dagger \hat{a}_r \hat{a}_s}
\end{align}
where $h_{pq}$ are the coefficients and $\hat{a}_p^\dagger$ and $\hat{a}_q$ are creation and annihilation operators for electron in molecular orbitals $p$ and $q$, respectively.
where $h_{pqrs}$ are the two-electron interaction coefficients, and $\hat{a}_p^\dagger, \hat{a}_q^\dagger, \hat{a}_r, \hat{a}_s$ are creation and annihilation operators for electrons in molecular orbitals $p, q, r, s$, respectively.

\begin{figure}
    \centering
    \begin{quantikz}
    \lstick[wires=5]{$\ket{\psi_{in}}$} & \gate{R_y(\theta_{0})} & \qw & \ctrl{1} & \qw \\
    & \gate{R_y(\theta_{1})} & \qw & \targ{} & \qw \\
    & \vdots  \\
    & \gate{R_y(\theta_{n-1})} & \qw & \ctrl{1} & \qw \\
    & \gate{R_y(\theta_{n})} & \qw & \targ{} & \qw \\
    \end{quantikz}
    \caption{One layer of hardware-efficient ansatz. The single-qubit parameterized gates are inserted on all qubits and two-qubit parameterized gatesa are inserted on all available connections.}
    \label{fig:HEA}
\end{figure}
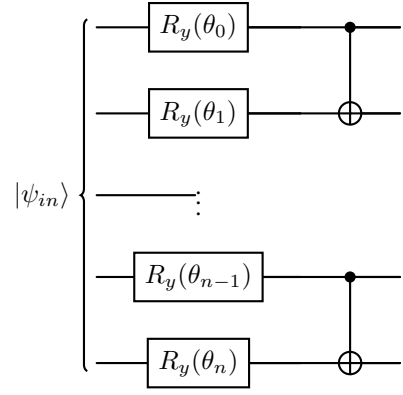

One typical ansatz is the hardware-efficient ansatz, the unitary matrix can be represented :
\begin{align}
  U(\vec{\theta}, \vec{\phi}) &= \bigotimes_{i=1}^{n} R(\theta_i, \phi_i) \cdot \prod_{\langle i,j \rangle \in E} CZ_{i,j} \\
  \ket{\psi(\vec{\theta}, \vec{\phi})} &= U^{(L)}(\vec{\theta}^{(L)}, \vec{\phi}^{(L)}) \cdots U^{(1)}(\vec{\theta}^{(1)}, \vec{\phi}^{(1)}) \ket{0}^{\otimes n} \\
  C(\vec{\theta}, \vec{\phi}) &= \langle \psi(\vec{\theta}, \vec{\phi}) | H | \psi(\vec{\theta}, \vec{\phi}) \rangle
\end{align}
Here, $n$ represents the number of qubits, $\vec{\theta}$ and $\vec{\phi}$ are vectors of the parameters for the rotation gates, and $E$ represents the set of edges for the entangling gates. $H$ is the Hamiltonian of the system being studied and $C(\vec{\theta}, \vec{\phi})$ is now the objective function to minimize.
An example of hardware-efficient ansatz is given in Figure~\ref{fig:HEA}, we can see that the single-qubit parameterized gates are inserted on all qubits and two-qubit parameterized gatesa are inserted on all available connections. Figure~\ref{fig:HEA} demonstrates only one layer of hardware-efficient ansatz, while multiple layers are used in real VQE tasks.

Another typical ansatz is the UCCSD ansatz, a chemistry-inspired ansatz.
First, we need to define a single excitation operator ($\hat{T}_1$), and a double excitation operator ($\hat{T}_2$):
\begin{align}
\hat{T}_1 &= \sum_{i=1}^{N{occ}} \sum_{a=1}^{N_{virt}} t_{i}^{a} \hat{a}^{\dagger}_{a} \hat{a}_{i} \\
\hat{T}_2 &= \sum_{i,j=1}^{N{occ}} \sum_{a,b=1}^{N_{virt}} t_{ij}^{ab} \hat{a}^{\dagger}_{a} \hat{a}^{\dagger}_{b} \hat{a}_{j} \hat{a}_{i}
\end{align}
Here, $N_{occ}$ and $N_{virt}$ represent the number of occupied and virtual orbitals, respectively, and $t_{ia}$ are the single excitation amplitudes and $t_{ij}^{ab}$ are the double excitation amplitudes. $\hat{a}^{\dagger}{a}$ and $\hat{a}{i}$ are the creation and annihilation operators for the respective orbitals.
Then we have the total excitation operator ($\hat{T}$) and the UCCSD unitary operator ($\hat{U}$):
\begin{align}
\hat{T} &= \hat{T}_1 + \hat{T}_2\\
\hat{U} &= e^{\hat{T} - \hat{T}^{\dagger}}
\end{align}
Afther the UCCSD ansatz is applied to a reference state $\ket{\Phi_0}$, we again have the objective function $C(t_{ia}, t_{ij}^{ab})$ to minimize.
\begin{align}
\ket{\psi_{UCCSD}} &= \hat{U} \ket{\Phi_0} \\
C(t_{i}^{a}, t_{ij}^{ab}) &= \langle \psi_{UCCSD} | H | \psi_{UCCSD} \rangle
\end{align}

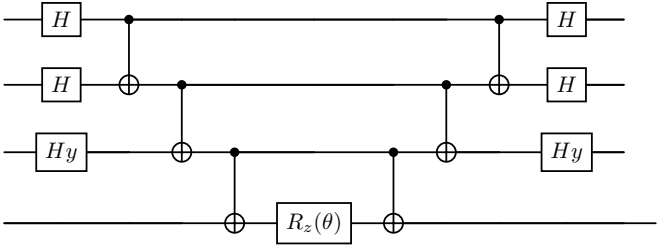
\begin{figure}
    \centering
    \scalebox{0.85}{
        \begin{quantikz}
        \qw&\gate{H} &\ctrl{1}&\qw&\qw&\qw&\qw&\qw&\ctrl{1}&\gate{H} &\qw\\
        \qw&\gate{H} &\targ{}&\ctrl{1}&\qw&\qw&\qw&\ctrl{1}&\targ{}&\gate{H} &\qw\\
        \qw&\gate{Hy}&\qw&\targ{}&\ctrl{1}&\qw&\ctrl{1}&\targ{}&\qw&\gate{Hy} &\qw\\
        \qw&\qw&\qw&\qw&\targ{}&\gate{R_z(\theta)}&\targ{}&\qw&\qw&\qw&\qw& \\
        \end{quantikz}
    }

    \caption{Example of UCC ansatz, we can see from the figure that the propotion of parameterized gates is very limited. We may introduce more flexibility into the ansatz by adding more parameterized gates into it.}
    \label{fig:UCC}
\end{figure}

Figure~\ref{fig:UCC} demonstrates a simple example of UCC ansatz, we can see that the circuits have multiple gates with only one parameterized $R_z(\theta)$ gate. The UCCSD ansatz has been considered as the gold standard for the design of ansatz circuits.

\subsection{Diffusion model}
Diffusion model is a machine learning model adapted from the  diffusion probabilistic models, first introduced by Jascha Sohl-Dickstein in 2015 ~\cite{DBLP:journals/corr/Sohl-DicksteinW15}. 
It is a type of generative model designed to remove Gaussian noise added to a graph while maintaining the graph's structure. This model has demonstrated its ability to preserve the underlying organization of the graph~\cite{DBLP:journals/corr/abs-2006-11239,DBLP:journals/corr/abs-2009-05475,DBLP:journals/corr/abs-2011-13456}. In this paper, we show that this characteristic can also be adopted to produce high-performance ansatz.
Based on the scheme developed by Jonathan Ho~\cite{DBLP:journals/corr/abs-2006-11239}, the training process for the diffusion model comprises two stages. 
The initial stage involves the gradual addition of Gaussian noise to an image, referred to as the forward process. 
Subsequently, the second stage (backward process) trains the parameters, enabling the model to learn noise reversal. 
For evaluation, the model employs parameters that are derived from the backward process on white noise, resulting in a new graph.

\begin{algorithm}
\caption{Jonathan Ho's Algorithm for training~\cite{DBLP:journals/corr/abs-2006-11239}}
\label{alg:1}
\begin{algorithmic}[1]
\REPEAT
\STATE $x_0 \sim q(x_0)$
\STATE $t \sim$ Uniform$({1,...,T})$
\STATE $\epsilon \sim N(0,I)$
\STATE Take gradient descent step on $$\nabla _\theta \| \epsilon \sim \epsilon _\theta (\sqrt{\overline{\alpha}_t} x_0 + \sqrt{1 - \overline{\alpha}_t} \epsilon, t)\|^2$$
\UNTIL converged
\end{algorithmic}
\end{algorithm}

\begin{algorithm}
\caption{Jonathan Ho's Algorithm for sampling~\cite{DBLP:journals/corr/abs-2006-11239}}
\label{alg:2}
\begin{algorithmic}[1]
\STATE $x_T \sim N(0,I)$
\FOR{$t = T,...,1$}
\STATE $z \sim N(0,I)$ if $t > 1$, else $z = 0$
\STATE $x_{t-1} = \frac{1}{\sqrt{\overline{\alpha}_t}}(x_t - \frac{1 - \alpha _t}{\sqrt{1 - \overline{\alpha}_t}}\epsilon _\theta(x_t,t)) + \sigma _tz$
\ENDFOR
\RETURN $x_0$
\end{algorithmic}
\end{algorithm}

As shown in the algorithm~\ref{alg:1} and the algorithm~\ref{alg:2}, the process of diffusion model can be divided into the training part and the testing part. In the training part, for the forward process, we add Gaussian noise to the graph using a Markov Chain: $$q(x_{t-1}| x_t) = N(x_t;\sqrt{1-\beta _t} x_{t-1}, \beta _t I)$$
with $x_t$ representing the image's status at time t, and $q(x_{t-1}|x_t)$ representing the transition from $x_{t-1}$ to $x_t$. And $\beta _t$ represents the coefficient of the noise we are adding at time t. Using the notation $\alpha _t = 1 - \beta _t$ and $\overline{\alpha}_t = \Pi \alpha _t$, the equation becomes:
$$q(x_{t-1}| x_t) = N(x_t;\sqrt{\overline{\alpha}_t} x_{0}, (1 -\overline{\alpha}_t) I)$$
Thus a state transition of a graph at any time t can be simply represented by an equation in terms of the original state of the graph. In this experiment specifically, We chose a linear schedule of $\beta$ from $10^{-4}$ to $0.01$.
For the backward process, we simply take gradient descent step on $\nabla _\theta \| \epsilon \sim \epsilon _\theta (\sqrt{\overline{\alpha}_t} x_0 + \sqrt{1 - \overline{\alpha}_t} \epsilon, t)\|^2$, with $\epsilon$ being the result of parameterization: $x_t(x_0,\epsilon) = \sqrt{\overline{\alpha}_t}x_0 + \sqrt{1-\overline{\alpha}_t}\epsilon$.
We used a 2-D Unet for our training process, which consists of two downsampling layers and two upsampling layers based on ~\cite{DBLP:journals/corr/RonnebergerFB15}.

In the sampling process, the diffusion model takes a random noise as a noisy image after timestep $T$, and reverse the forward process using the trained parameters at each timestep, until the assumed ``original image'' is achieved. In our implementation, we take $\sigma _t$ as a result of parameterization $\sigma _t^2 = \beta _t$


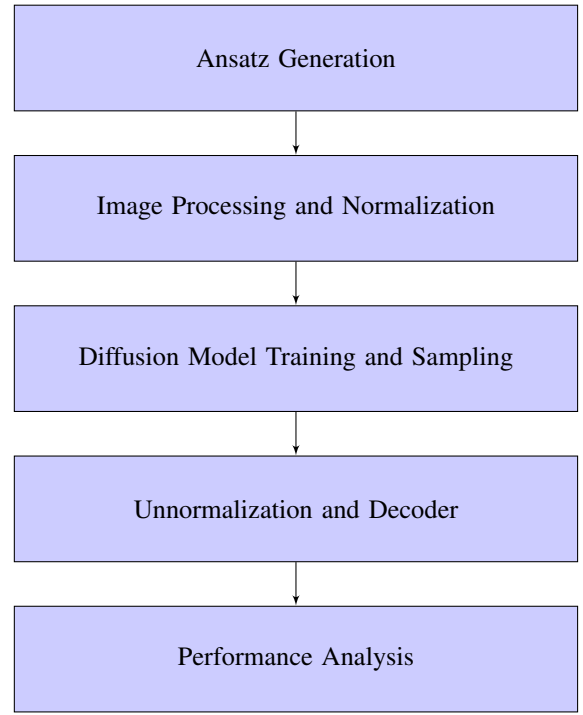
\begin{figure}[htbp]
\centering
    \begin{tikzpicture}[node distance=2cm,scale=20]
    
    \tikzstyle{block} = [rectangle, draw, fill=blue!20, 
        text width=0.4\textwidth, text centered, minimum height=4em]
    \tikzstyle{line} = [draw, -latex']
        
    \node [block] (ansatz) {Ansatz Generation};
    \node [block, below of=ansatz] (image) {Image Processing and Normalization};
    \node [block, below of=image] (train) {Diffusion Model Training and Sampling};
    \node [block, below of=train] (unnormalize) {Unnormalization and Decoder};
    \node [block, below of=unnormalize] (performance) {Performance Analysis};
    
    \path [line] (ansatz) -- (image);
    \path [line] (image) -- (train);
    \path [line] (train) -- (unnormalize);
    \path [line] (unnormalize) -- (performance);
    
    \end{tikzpicture}
\caption{In our proposed technique's workflow, we begin by generating ansatz using the Unitary Coupled Cluster (UCC) method, which serves as our dataset. Next, we convert the ansatz circuits into images suitable for image processing and normalization. Afterward, the dataset is input into the diffusion model, and we collect the samples. These samples correspond to ansatz circuits that exhibit similar structures to the input ansatz. Finally, we evaluate these ansatz using the Variational Quantum Eigensolver (VQE) tasks.}
\label{fig:workflow}
\end{figure}

\section{Methodology}


\subsection{Dataset Generation}
In the first step, we need to generate a group of ansatz that will later be transformed into images.
We generate the UCC ansatz from a random Pauli string with the scheme developed in ~\cite{Whitfield_2011}. 
When we encouter a Pauli string like XXYZ, we add Hadamard gates onto the first two qubits and $H_y$ gates onto the next qubit. We don't need to add extra gate for the remaining ``Z''. Then we connect the qubits with CNOT gates and insert a $R_z(\theta)$ gate in the middle as the parameterized gate. 
The next step is to reverse the aforementioned Hadamard gates and CNOT gates to generate a symmetric form as demonstrated in Figure~\ref{fig:UCC}

\begin{figure}
\begin{subfigure}{.15\textwidth}
  \centering
  \includegraphics[width=.8\linewidth]{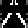}
  \label{fig:sfig1}
\end{subfigure}%
\begin{subfigure}{.15\textwidth}
  \centering
  \includegraphics[width=.8\linewidth]{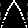}
  \label{fig:sfig2}
\end{subfigure}%
\begin{subfigure}{.15\textwidth}
  \centering
  \includegraphics[width=.8\linewidth]{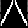}
  \label{fig:sfig2}
\end{subfigure}
\caption{Examples of the images that are generated from our UCC ansatz circuits.}
\label{fig:exampe_image}
\end{figure}

Then we need to transform the generated ansatz circuits into images that can be handled by the diffusion model. In this paper, we propose to use different values of pixels to represent the different gates inside the ansatz circuits. In this way, we generate a groups of images that correspond with the group of UCC ansatz. The example is given in Figure~\ref{fig:exampe_image}, we can tell from the figures that UCC ansatz usually preservers a ``V'' shape. For each figure, the background is set to zero, which is black in images.
For different numbers of qubits, we first generate random Pauli strings and their associated ansatz circuits. Next, these circuits are converted into images and undergo a normalization process, in which the images are resized to 28x28 dimensions. For each qubit number, we produce 10,000 images.

\subsection{Image decoder}

\begin{figure}
\begin{subfigure}{.15\textwidth}
  \centering
  \includegraphics[width=.8\linewidth]{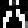}
  \label{fig:sfig1}
\end{subfigure}%
\begin{subfigure}{.15\textwidth}
  \centering
  \includegraphics[width=.8\linewidth]{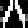}
  \label{fig:sfig2}
\end{subfigure}%
\begin{subfigure}{.15\textwidth}
  \centering
  \includegraphics[width=.8\linewidth]{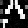}
  \label{fig:sfig2}
\end{subfigure}
\caption{Examples of the sampled images that are generated from the diffusion model. }
\label{fig:output_image}
\end{figure}

\begin{figure}
    \centering
    \includegraphics[width=0.5\textwidth]{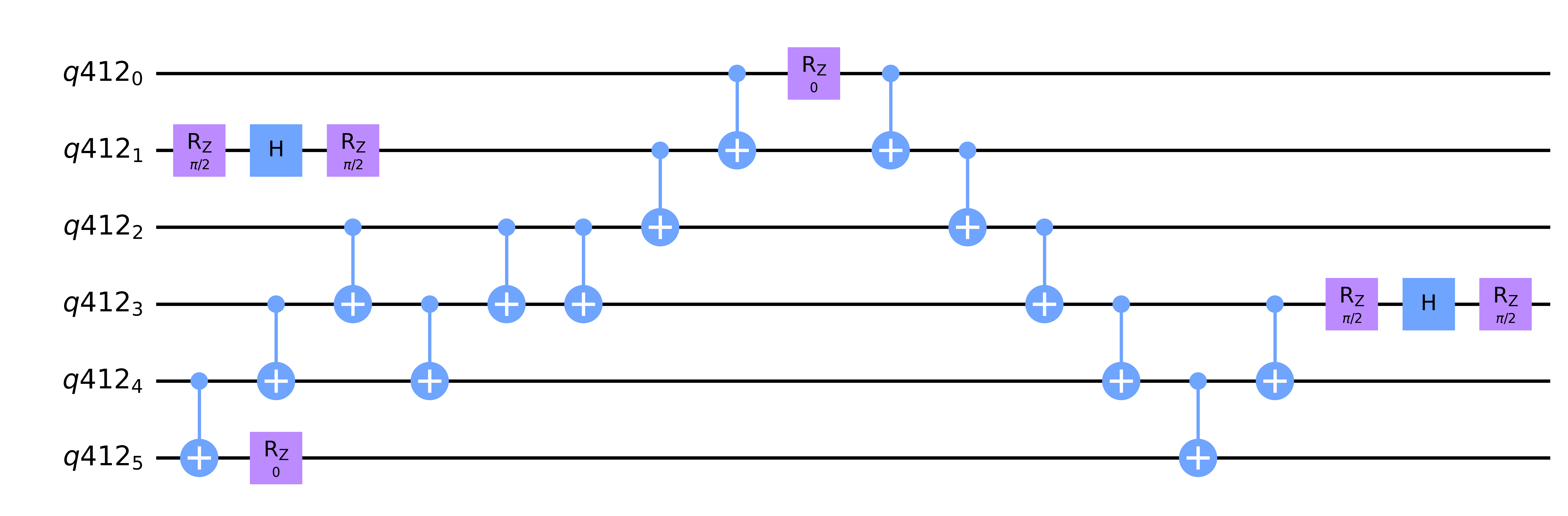}
    \caption{Example of quantum circuits from decoder.}
    \label{fig:decoded_circ}
\end{figure}

Upon acquiring the image dataset derived from the generated UCC ansatz, we feed it into the diffusion model. We then obtain samples from the diffusion model, with the sampled images maintaining the structure of the input dataset, as illustrated in the Figure~\ref{fig:output_image}. The subsequent step involves decoding these images into quantum circuits, which can serve as ansatz in the VQE task.
For the decoder, we first calculate the height and width required for the desired number of qubits, so that each gate can be represented by exactly one pixel. Given the ansatz generation scheme, the height of the graph should always be the number of qubits $N$, and the width should always be twice the number of non-identity layers plus one. The fraction of identity layers with respect to the whole circuit of the generated image is approximated according to the fraction of the black pixels of each line. Then, we interpret the graph pixel by pixel until we have the entire ansatz. This allows us to accurately represent each gate in the ansatz and ensure that the decoder can properly interpret the image.
The encoder and decoder is designed in a way that can best distinguish different types of gates as illustrated in the Table~\ref{tab:encod}.
Figure~\ref{fig:decoded_circ} illustrates a quantum circuit following the decoding step. However, this quantum circuit does not guarantee optimal performance in VQE tasks. Consequently, we generate a set of ansatz candidates and assess their performance on VQE tasks.

\begin{figure}
    \centering
    \includegraphics[width=0.5\textwidth]{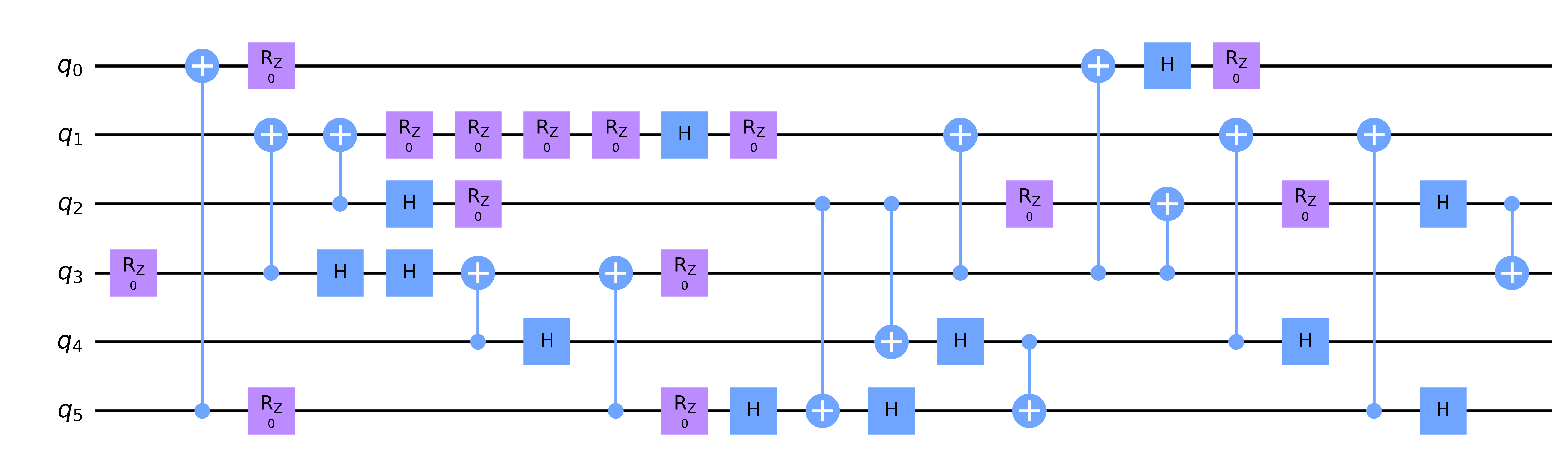}
    \caption{Example of randomly generated circuits.}
    \label{fig:random}
\end{figure}

\subsection{VQE evaluation}

\begin{table}
  \centering
  \begin{tabular}{>{\raggedright\arraybackslash}m{5em} >{\centering\arraybackslash}m{3cm} >{\centering\arraybackslash}m{3cm}}
    \toprule
    Gate types           & Encoder & Decoder                  \\
    \midrule
    Background           & 0       & $0 \sim 30$                   \\
    \addlinespace
    $R_z(\theta)$        & 40      & $30 \sim 50$                   \\
    \addlinespace
    $H$                  & 80      & $70 \sim 90$                   \\
    \addlinespace
    $Hy$                 & 100     & $90 \sim 120$                  \\
    \addlinespace
    $CX$                 & 255     & $160 \sim 255$                 \\
    \addlinespace
    Decoded as background & N/A & $50 \sim 70, 120 \sim 160$ \\
    \bottomrule
  \end{tabular}
  \caption{Encoding and decoding policy}
  \label{tab:encod}
\end{table}

After acquiring the images from the diffusion model and decoding them into quantum circuits, we evaluate the performance of the generated ansatz circuits using VQE tasks for the molecules $H_2,\ LiH\ and\ H_2O$.
We adopt the framework TorchQuantum~\cite{wang2022quantumnas} to evaluate the performance of these ansatz circuits. 
The Hamiltonians of the molecules are obtained from Qiskit~\cite{cross2018ibm}.
We also generate random ansatz for reference
We evaluate the results by comparing them with the randomly produced circuits having an equal number of qubits and twice the gate count. 
The Figure~\ref{fig:random} presents an example of a randomly generated ansatz.
The findings reveal that the minimum energy obtained through random circuits significantly deviates from the one achieved by our devised ansatz.

\begin{figure}[htbp]

    \centering
    
    \begin{subfigure}[b]{0.5\textwidth}
        \centering
        \includegraphics[width=\textwidth]{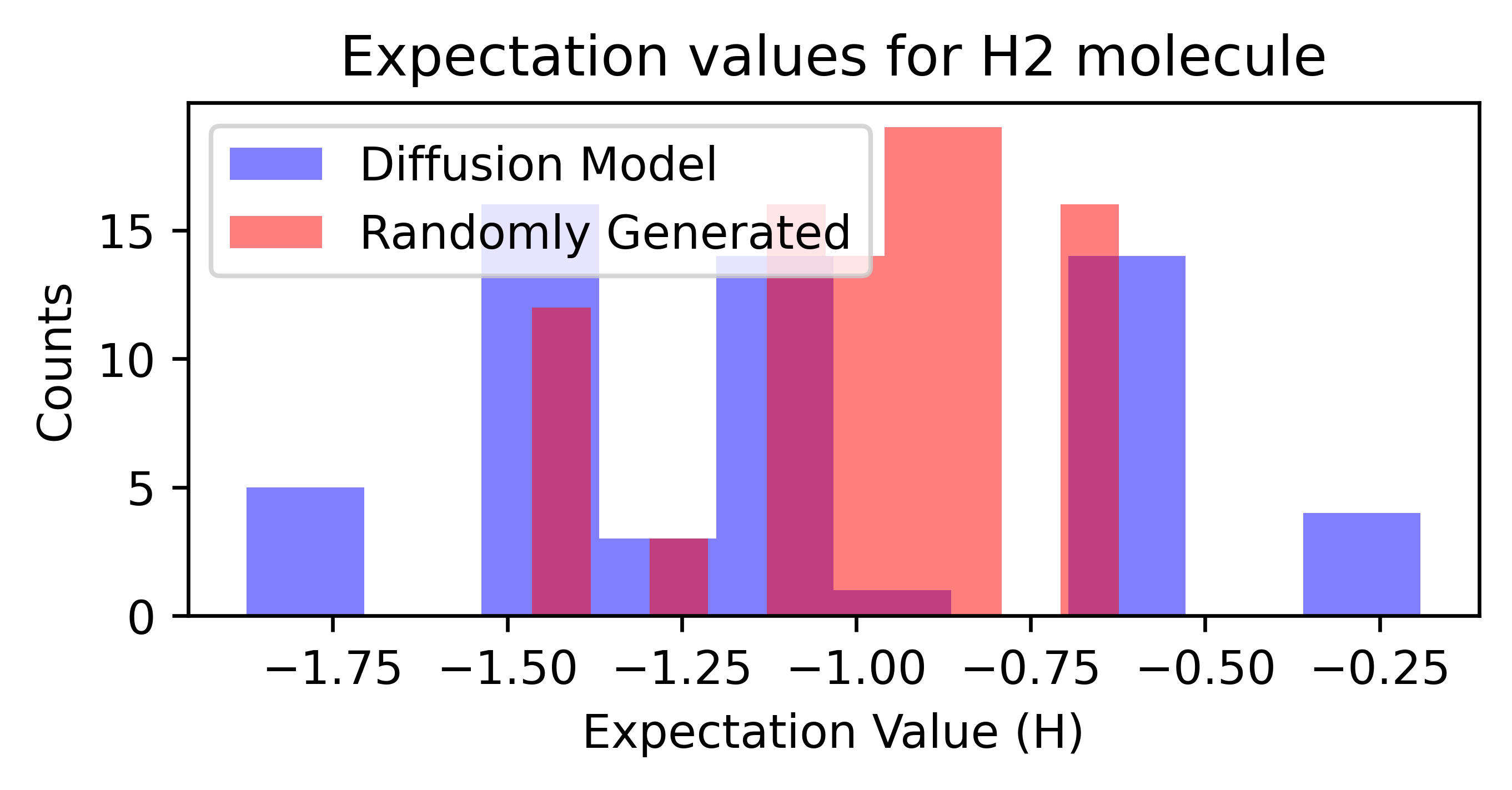}
        \label{fig:image-a}
    \end{subfigure}
    
    \vspace{-0.5cm} 
    
    \begin{subfigure}[b]{0.5\textwidth}
        \centering
        \includegraphics[width=\textwidth]{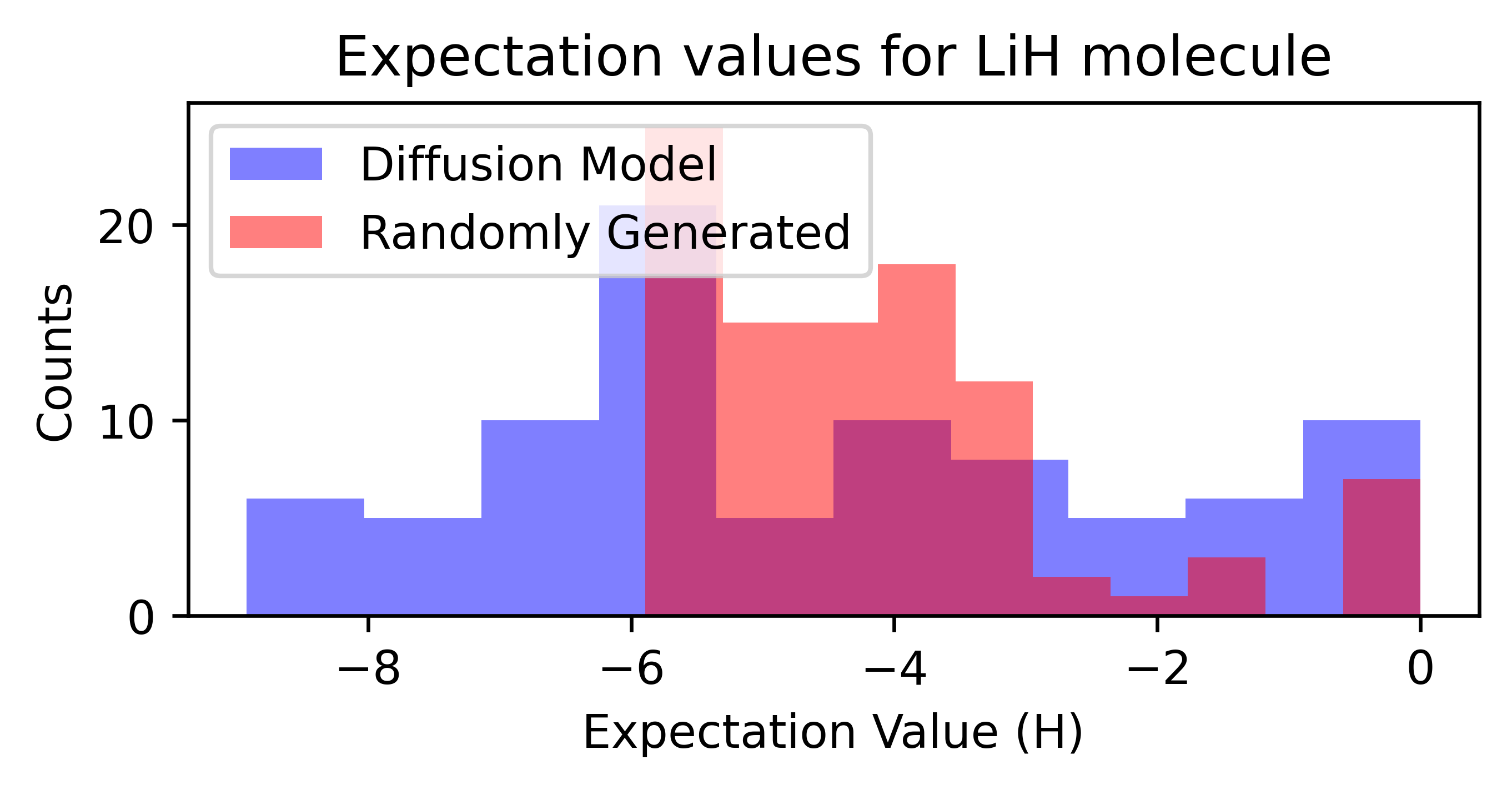}
        \label{fig:image-b}
    \end{subfigure}
    
    \vspace{-0.5cm} 
    
    \begin{subfigure}[b]{0.5\textwidth}
        \centering
        \includegraphics[width=\textwidth]{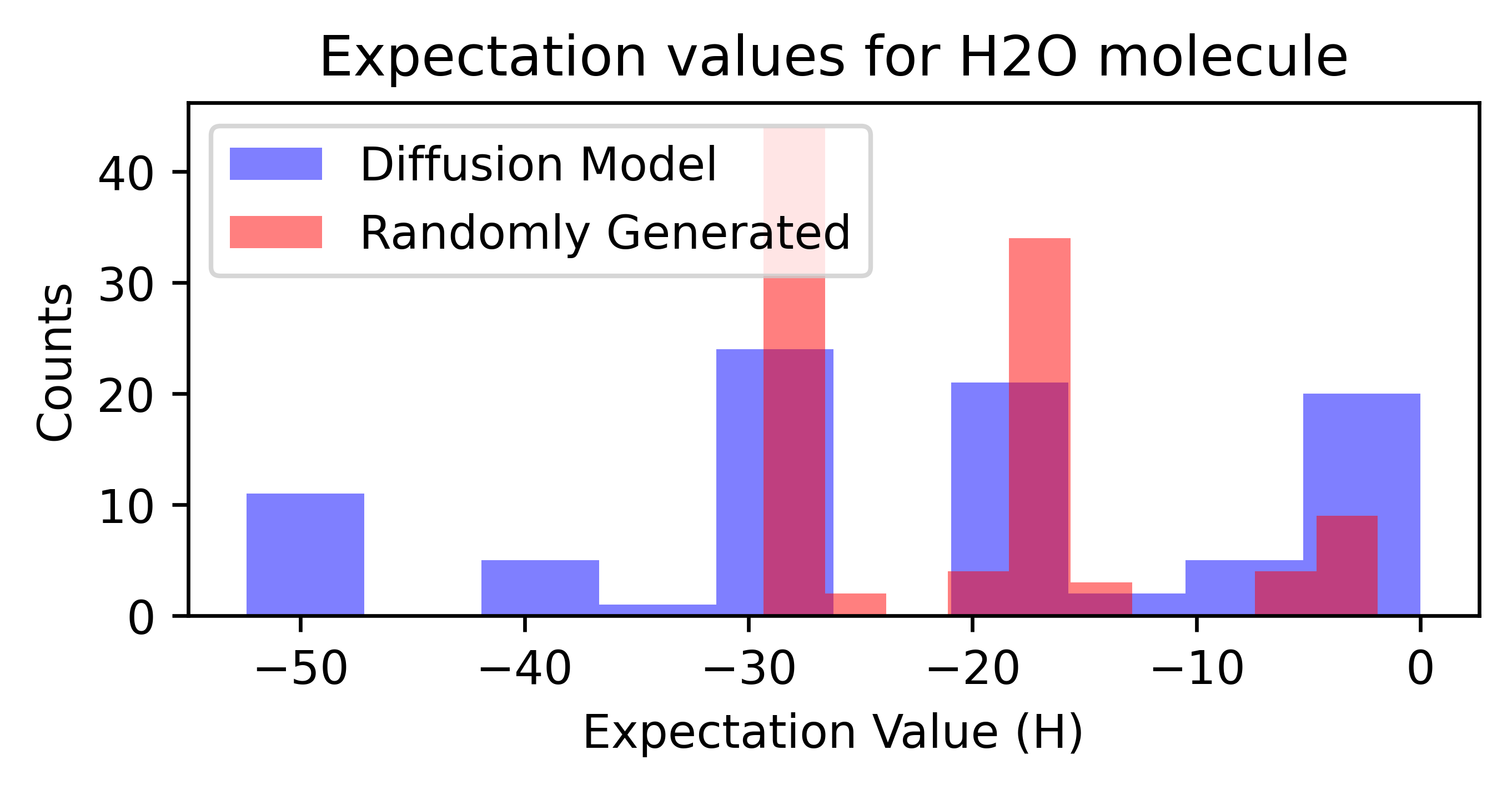}
        \label{fig:image-c}
    \end{subfigure}
    
    \caption{The expectation values obtained from different ansatz types indicate that those from the diffusion model are more widely spread and, most importantly, can reach the optimal ground state energy of the molecules. In contrast, the randomly generated ansatz is unable of determining the accurate energy of the molecules. The diffusion model-generated ansatz eliminates the need for multi-layer UCCSD ansatz, as it already includes sufficient parameterized gates to guarantee the expressibiltiy of the ansatz circuits.}
    \label{fig:three-figures}
    
\end{figure}

\section{Results}

\subsection{Experimental setup}
Our ansatz is assessed within the TorchQuantum framework on a server equipped with dual Xeon E5-2630 v3 CPUs and 64 GB of RAM. The diffusion model's training and sampling are conducted on a server with an NVIDIA Tesla K40m GPU. For VQE task training, we employ the ADAM optimizer, set a maximum of 100 iterations with a learning rate of 0.1.

\subsection{Performance of generated ansatz}

As shown in the Figure~\ref{fig:three-figures}, for the VQE tasks of molecules $H_2$, $LiH$, and $H_2O$, the ansatz generated by the diffusion model yields expectation values of -1.873, -8.921, and -52.396, respectively. Meanwhile, the randomly generated ansatz provides expectation values of -1.464, -5.944, and -29.472, respectively. As a result, the diffusion model-generated ansatz returns superior results by preserving the UCC ansatz structures. Such accuracy is achieved without the need for the NAS process proposed in ~\cite{wang2022quantumnas}, which introduces significant computational overhead.

\section{Future work}
Large language models, such as ChatGPT~\cite{lund2023chatting}, have had a significant impact recently. There is potential for the advancements in artificial intelligence to further benefit quantum computing research. 
It is expected that we will see an increase in the use of AI in quantum computing research. 
In this paper, we explores the use of AI models in the design of ansatz circuits for variational quantum algorithms.
Our ansatz generator adopts diffusion model and creats ansatz circuits that preserve certain structures of the UCC ansatz. 
The efficacy of the generated ansatz can be further tested on noisy simulators and NISQ devices. 
In our technique, the encoder and decoder can be optimized to ensure that the generated circuits possess desirable features. 
It is also possible to further optimize the generation of ansatz circuits by considering the underlying hardware topology of the quantum devices.

\section{Conclusion}
The aim of this paper is to introduce the use of a diffusion model in generating ansatz circuits for the variational quantum eigensolver. Our objective is to keep certain structures from the UCC ansatz, while simultaneously inserting additional parameterized gates into the ansatz circuits. To achieve this, we first prepare a substantial set of random UCC ansatz and convert them into images that can be processed by the diffusion model. The diffusion model is then trained and sampled. The sampled images are then decoded into quantum circuits, which serve as our ansatz candidates. We assess the performance of these ansatz circuits on VQE tasks, and demonstrate that they exhibit superior accuracy when compared to random ansatz circuits of larger size.

\printbibliography
\end{document}